\documentclass[12pt]{article}

\usepackage{amsmath,amsthm, amsfonts, amssymb, amsxtra, amsopn}
\usepackage{pgfplots}
\pgfplotsset{compat=1.13}
\usepackage{pgfplotstable}
\usepackage{graphicx,grffile}
\usepackage{multirow}
\usepackage{booktabs}
\usepackage{listings}
\usepackage{cmap}
\usepackage{colortbl}
\usepackage{adjustbox}
\usepackage{epsfig}
\usepackage{bold-extra}

\usepackage[tableposition=top,font=small,skip=5pt]{caption}
\usepackage{subcaption}
\usepackage{makecell} 

\PassOptionsToPackage{hyphens}{url}

\usepackage{hyperref}
\hypersetup{colorlinks=true,linkcolor=black,citecolor=black,urlcolor=blue,filecolor=black}
\hypersetup{pdfpagemode=UseNone,pdfstartview=}

\usepackage{enumitem}
\setlist[itemize]{noitemsep, topsep=0pt}

\advance\oddsidemargin by -0.35in
\advance\textwidth by 0.7in

\advance\topmargin by -0.4in
\advance\textheight by 0.8in

\long\def\symbolfootnotetext[#1]#2{\begingroup%
\def\thefootnote{\fnsymbol{footnote}}\footnotetext[#1]{#2}\endgroup}

\DeclareMathOperator{\thth}{th}
\DeclareMathOperator{\ndnd}{nd}

\let\vvv=\v
\def\v{{\tt v}}

\title{Steganographic Capacity of Deep Learning Models}

\author{Lei Zhang\footnotemark[1]\ \ \ \ 
Dong Li\footnotemark[2]\ \ \ \ 
Olha  Jure\vvv{c}kov\'{a}\footnotemark[3]\ \ \ \ 
Mark Stamp\footnotemark[1]\,\,\footnotemark[4]}

\begin{document}

\symbolfootnotetext[1]{Department of Computer Science, San Jose State University}
\symbolfootnotetext[2]{Shanghai AI Laboratory}
\symbolfootnotetext[3]{Faculty of Information Technology, Czech Technical University in Prague}
\symbolfootnotetext[4]{mark.stamp$@$sjsu.edu}

\maketitle

\abstract
As machine learning and deep learning models become ubiquitous, it is inevitable that there will be 
attempts to exploit such models in various attack scenarios. For example, in a steganographic-based attack, 
information could be hidden in a learning model, which might then be used to distribute malware, or for 
other malicious purposes. In this research, we consider the steganographic capacity of several 
learning models. Specifically, we train a Multilayer Perceptron (MLP), 
Convolutional Neural Network (CNN), and Transformer model on a challenging malware classification 
problem. For each of the resulting models, we determine the number of low-order bits of the trained 
parameters that can be altered without significantly affecting the performance of the model. We find 
that the steganographic capacity of the learning models tested is surprisingly high, and that in each case, 
there is a clear threshold after which model performance rapidly degrades.

\section{Introduction}

Steganography, or information hiding, consists of embedding information 
in another message or physical object~\cite{Mohammed_2021}. While 
cryptography also hides information, it does so by converting the information into a human 
unreadable form~\cite{Whitfield_2022}. The main difference between these two techniques is 
that cryptography alters the structure of the secret information but does not hide the fact
that communication is taking place, while steganography hides the information
in another medium that is not intended for such communication~\cite{Rina_2015}. 
Modern steganographic techniques have been developed for a wide range of
data types, including text, images, audio, video, 
and even networking data~\cite{Agarwal_2013}. 

Machine learning (ML), which can be considered as a subfield of artificial intelligence, 
enables computers to learn important information from training data~\cite{Stamp_2022}. 
Today, ML models are widely used to deal with a vast array of problems, including speech recognition, 
image recognition, sentiment analysis, language translation, malware detection, and so on, with new applications 
being constantly developed. Deep learning (DL) models are the subset of ML models that are based  
on neural networking techniques.

Machine learning models are a plausible cover media in steganography for the following reasons.
\begin{enumerate} 
\item Machine learning models are rapidly becoming ubiquitous.
For example voice-activated search assistants were used by approximately~3.25 billion people 
worldwide in~2021, out of a world population of~7.9 billion~\cite{ML_Stat}.
\item It is likely that the information hiding capacity of most machine learning models is substantial. 
Machine learning models typically include a large number of weights or other trained parameters,
and it is known that learning models typically do not require high precision in their trained 
parameters. For example, the most popular algorithm used to train Support Vector Machines (SVM) 
relies on the fact that limited precision is sufficient~\cite{Stamp_2022}. 
As another example, in neural networking-based models,
many---if not most---neurons tend to atrophy during training, and such weights contribute little to the 
trained model. By relying on such redundant neurons,
the authors of~\cite{EvilModel} show that they can hide~36.9MB of malware
within a~178MB AlexNet architecture, with only a~1\%\ degradation in performance. These 
changes did not affect the structure of the model and the embedded malware was
not detected by any of the anti-virus systems tested.
\item Machine learning models may be an ideal cover media for malicious attacks. For example, 
as in~\cite{EvilModel}, malware could be embedded in a learning model.
It is even conceivable that a specific predetermined input to the 
model could be used to trigger an embedded malware-based attack.
\end{enumerate}

In this research, we focus on the fact that, in general, learning models do not require 
high precision in their trained parameters. Therefore,
as a measure of the inherent steganographic capacity of learning models, we determine
the number of low-order bits in each weight that can be used for information hiding purposes. 
We embed information in the~$n$ low-order bits of the weights of trained models, 
and graph the model accuracy as a function of~$n$.
We analyze three DL models: Multi-Layer Perceptron (MLP), 
Convolutional Neural Network (CNN), and a Transformer architecture. 
We train and test each of these models on a dataset that 
contains~10 different malware families, with a total of~15,356 samples. 

The remainder of the paper is organized as follows.
Section~\ref{chap:2} gives relevant background information on steganographic techniques
and the various machine learning models used in this research. 
Section~\ref{chap:3} provides details on the dataset employed in our experiments, 
along with a high-level view of our experimental design. Our results are
presented and discussed in Section~\ref{chap:4}.
Finally, Section~\ref{chap:5} gives our conclusions, as well as outlining
potential avenues for further research.

\section{Background}\label{chap:2}

In this section, we discuss several relevant background topics. First, we consider steganography,
then we introduce the learning models that are used in this research. We 
conclude this section with a discussion of related work.

\subsection{Steganography}

The word ``steganography'' is a combination of two Greek roots: \textit{stegan\'{o}s}, which 
means ``concealed or hidden'', and \textit{graphein}, which translates as ``drawing or writing''~\cite{Fiscutean_2021}. 
Thus, steganography is the art and science of embedding secret 
information inside unremarkable cover media that does not raise suspicions~\cite{Stanger_2020}. 
In modern practice, steganography consists of concealing information or messages within seemingly 
innocuous data or media, such as images, audio, video, or network communication, 
among many other possibilities~\cite{Agarwal_2013}. 

Steganography involves embedding secret data 
into a cover media in a way that is imperceptible to human senses and difficult to detect without specialized 
tools and knowledge. Such techniques have been used throughout history for various purposes, 
including espionage, communication in times of war, and digital watermarking. 
With the advancement of digital technology, 
steganography has found applications in modern information security, digital forensics, 
and multimedia communications, among others. It is an evolving field with ongoing research 
and development of new techniques to enhance its security and application in various domains.

Cryptography protects a secret message by transforming it into an unintelligible format to hide 
the meaning of the message, while steganography aims to hide the presence of the 
original message~\cite{Stamp_2021}. Steganography dates at least
to ancient Greece and, in fact, it predates cryptography as a means 
of secret communication~\cite{Stamp_2021}. 
An historical example of steganography was the use
of invisible ink during the American Revolutionary War 
to pass messages. As another example, during World War~II, photosensitive glass~\cite{John_2017} and 
microdots~\cite{Clelland_1999} were used to embed information in other messages. 
Today, hiding information in image files on computing systems is the most common method 
of steganography. 

A textbook example of a modern steganographic application consists of hiding information in the low order
RGB bits of an uncompressed image file, such as a \texttt{bmp} image~\cite{Stamp_2021}. 
Since the RGB color scheme
uses a byte for each of the R (red), G (green), and B (blue) color components, there 
are~$2^{24} > 16{,}000{,}000$ colors available. However, many of the colors are indistinguishable to
the human eye, and hence there are a large number of redundant bits in an uncompressed image. 
The low-order RGB bits of each byte can be used to hide information
in such an image, without changing the carrier image in any perceptible way. Provided that the 
intended recipient knows which image is used for hiding information, and knows how to extract
the information, communication can take place between a sender and receiver, without
it being apparent that such communication has even occurred. The steganographic capacity of
an uncompressed image file is surprisingly large; for example, in~\cite[Section~5.9.3]{Stamp_2021}
it is shown that the entire \textit{Alice's Adventures in Wonderland} book can be hidden in the 
low order RGB bits of an image of Alice from the \textit{Alice} book itself.

The image-based steganographic system described in the previous paragraph is not robust, that is,
it is trivial to disrupt the communication, without the disruption affecting the non-steganographic
use of such images: If we suspect that the low-order RGB bits of \texttt{bmp} files are being
used for steganographic purposes, we can simply randomize the low-order bits of all \texttt{bmp} images.
For any such images that were being used for information hiding, the information would be lost, and
for any innocent images that were not used for information hiding, the image would not be affected
in any perceptible way. Much of the modern research into information hiding revolves around creating 
more robust steganographic techniques.

Steganography can be characterized by three important aspects, namely,
perceptual transparency, robustness, and capacity.  
\begin{itemize}
\item Perceptual transparency --- 
This refers to the ability of the steganographic process to hide the secret information
in a way that is imperceptible to human senses. This is a critical characteristic of steganography, 
which ensures that it is not obvious that the cover medium is being used for surreptitious communication. 
\item Robustness --- 
By definition, robustness is the ability to tolerate perturbations of a system without adversely affecting
its initial stable configuration~\cite{Wieland_2012}. In image-based steganographic techniques, the perturbations 
could be transformation, sharpening, filtering, scaling, cropping, and so on.
\item Capacity --- 
The amount of information that can be hidden in the cover medium is the capacity, which is related
to the practical redundancy in the cover media. The larger the capacity, 
the more information that can be hidden; equivalently, the smaller the cover medium that is needed. 
\end{itemize}
Achieving an optimal balance among 
these characteristics is a crucial consideration in the design and implementation of 
a steganographic technique, as it determines the effectiveness of the communications,
and the security of the concealed information.

In this research, we are interested in the steganographic capacity of machine learning 
models. Specifically, we hide information in the low-order bits of the weights of learning models.
While such a scheme is not robust, our work does provide an intuitive and practical means for information
hiding. We show that learning models have considerable redundancy, which
is the basis for more advanced steganographic techniques, with the analogy to uncompressed
image files being obvious.

\subsection{Learning Models}\label{sect:learningModels}

Machine learning (ML) and deep learning (DL) can be viewed as branches of 
artificial intelligence (AI). In general, ML refers to the use of statistical models and algorithms to enable machines 
to learn from data and improve their performance on a specific task. DL is 
the subset of machine learning that focuses on training Artificial Neural Networks (ANN)---generally 
with multiple hidden layers, which is the ``deep'' part of deep learning---to identify 
patterns and relationships in data. DL algorithms, 
which are designed to (loosely) mimic the structure and functioning of the human brain, 
have proven to be very effective in solving complex problems such as image and speech recognition, 
natural language processing, and even playing complex games. 

ML enables computers to learn important information, and improve from experience,
which saves humans from the work of extracting useful information from seemingly 
inscrutable data~\cite{Stamp_2022}. The process of machine learning begins with observations 
derived from datasets.  The primary goal of machine learning is to make computers learn with
minimal human intervention or assistance~\cite{Selig_2022}.

ML is applied in a wide and ever-growing range of important fields, 
including data security, finance, healthcare, fraud detection, and so on. In addition,
DL techniques have been used to successfully deal with such problems as speech recognition, 
image classification, sentiment analysis, and language translation, among many others~\cite{Duggal_2022}. 

Deep learning has gained significant attention and success in recent years due to 
its ability to automatically extract complex patterns and representations from raw data 
without extensive feature engineering. Through the process of training, deep learning models 
learn to recognize patterns, features, and relationships in data, enabling them to often perform tasks 
at a higher level than had previously been achieved using classic machine learning models.

Machine learning algorithms can be subdivided into three primary categories: 
supervised machine learning, unsupervised machine learning, and semi-supervised machine learning. 
Supervised machine learning uses labeled datasets to train the model.
Support Vector Machine, Multilayer Perceptron, $k$-Nearest Neighbors, Decision Trees, Random Forest, 
and Linear Regression are popular examples of supervised machine learning algorithms. 
In contrast, unsupervised machine learning techniques can be applied to unlabeled data. 
Clustering techniques, such as the well-known $K$-means algorithm,
are examples of unsupervised learning.
Semi-supervised machine learning can be viewed as a hybrid approach that combines
aspects of supervised and unsupervised algorithms.
In this paper, we only consider supervised learning techniques; specifically, we train
models to classify malware from several different families.

Next, we discuss each of the learning techniques that are employed in the experiments
in Section~\ref{chap:4}. Here, we introduce the DL techniques of Multilayer Perceptron, 
Convolutional Neural Networks, and Transformer models.

\subsubsection{Overview of Multilayer Perceptrons}

Multilayer Perceptrons (MLP) are a popular class of feedforward neural network architectures 
that are widely used for supervised learning tasks, including classification and 
regression~\cite{MLP_Intro}. MLPs consist of multiple layers of interconnected nodes, 
where each node receives input from the previous layer and produces output that is passed to 
the next layer.

The input layer of an MLP receives the input data, and the output layer produces the final prediction.
In between these layers, there can be one or more hidden layers that help to learn complex patterns 
in data. Each node in the hidden layers applies a nonlinear activation function to the weighted sum of 
its inputs, which helps to capture non-linear relationships in the data.

MLPs are trained using backpropagation, which is an optimization algorithm that adjusts 
the weights of the network based on the difference between the predicted output and the 
actual class label. The weights are updated using gradient descent, which iteratively adjusts 
the weights to minimize the error.

One of the main advantages of MLPs is their ability to learn complex patterns in the data, making 
them suitable for high-dimensional and non-linear datasets. However, MLPs can be computationally 
expensive to train, and they require a large amount of labeled data to achieve high accuracy.

\subsubsection{Overview of Convolutional Neural Networks}

Convolutional Neural Network (CNN) are one of the most popular DL techniques. 
CNNs were designed for efficient training on images, where local structure dominates,
but they have proven surprisingly useful for a wide range of problems---any problem domain
where local structure is most important is a good candidate for a CNN. The CNN 
architecture is composed of convolution layers, pooling layers, and a fully connected layer
(or, possibly, multiple fully connected layers).

A convolution layer performs a discrete convolutional operation on the output of the previous layer.
This can be viewed as applying a filter, where the parameters of the filter are learned.
The first convolutional layer is applied to the input data, and in the case of images
it learns basic structure, such as edges. Subsequent convolutional layers learn higher-level
and more abstract features.

The purpose of a pooling layer, which usually follows a convolution layer, is to reduce the dimensionality
of the problem, and thereby speed up the training process. Pooling may also serve to increase translation
invariance, which is highly desirable for image analysis.

\subsubsection{Overview of Transformer Models}

Transformers are a type of deep learning architecture that have revolutionized the field of 
natural language processing (NLP). Transformers were introduced 
in~\cite{Attention_Is_All_You_Need}, and are currently the state-of-the-art architecture 
for many NLP tasks, including machine translation, sentiment analysis, and question answering. 
They have also been successfully applied to other tasks, 
such as image classification and speech recognition.

The key innovation of Transformers is the self-attention mechanism, which allows the model 
to selectively attend to different parts of the input sequence when making predictions. 
All models use attention to some degree, but Transformer model in~\cite{Attention_Is_All_You_Need}
showed that explicit attention is far more powerful than had been previously realized.

Transformers consist of an encoder and a decoder module. The encoder takes an input sequence 
and generates a hidden-state representation that is designed to capture the meaning of the input. 
The decoder takes the hidden-state representation and generates the output one token at a time.

One of the key advantages of Transformers is their ability to handle variable-length input sequences 
without the need for padding or truncation. They also require less training time compared to traditional 
Recurrent Neural Networks, and can be parallelized more easily.

\subsection{Related Work}

As of this writing, the only closely related work that the authors are aware of
is~\cite{EvilModel}, in which a technique dubbed ``EvilModel'' is developed and analyzed. 
EvilModel hides malware inside of a neural network model. As an example, 
when a~36.9MB malware is embedded in a specific model, 
the accuracy of the model is only reduced by about~1\%. 
The authors of~\cite{EvilModel} embed the malware sample 
in a learning model by carefully selecting 
weights that have atrophied during training, and thus have little or no effect
on model performance. They then overwrite these entire weights.

The approach used in~\cite{EvilModel} is considerably different from the
experiments that we conduct in this paper. Here, we do not analyze models to find
insignificant weights, but instead we overwrite the least-significant bits of all 
weights in the various layers of a model.
Our approach is much simpler, in the sense that it requires no detailed analysis of the model,
yet we are able to embed signifcant amounts of data in all of the models that
we consider. Our results point to a more generic issue with respect to learning models,
namely, that a large steganographic capacity is inherent in models that
use a relatively large number of bits to store weights. 
As a byproduct of this research, we verify that only limited accuracy is required 
of the weights for the models that we consider.

\section{Implementation}\label{chap:3}

In this section, we first discuss the malware dataset used to train our learning models. 
Then we provide
details on the training of each of the models considered in this paper.
The steganographic capacity of these models is analyzed in Section~\ref{chap:4}, below.

\subsection{Dataset}

Malware families can be difficult to define precisely because they can vary in terms of their size, 
scope, and specific features. However, a family generally refers to a group of malware samples 
that have similarities in terms of their functionality, behavior, and code structure. Although the 
specific details of each sample may differ, members of a given family typically share a core code base 
that contains common functions, routines, and behaviors. This allows security researchers to identify 
and track specific malware families over time, even as the individual samples within the family continue 
to evolve and change. By analyzing these shared characteristics, researchers can develop more 
effective detection and mitigation strategies to protect against the threat of malware.

In this research, we consider a malware dataset from VirusShare~\cite{VirusShare}. 
This dataset contains more than~500,000 malware executables, which occupy more 
than~500GB of storage. Among the~500,000 malware executables, we have extracted the
top~10 families, in terms of the number of samples available per family. 
Specifically, we consider the malware families listed Table~\ref{tab:1}, 
which are given in descending order, based on the number of samples. 

\begin{table}[!htb]
\caption{Malware families}\label{tab:1}
\centering
\adjustbox{scale=0.85}{
\begin{tabular}{c|cc}\midrule\midrule
Family & Samples & Fraction of total\\ \midrule
\texttt{VBinject} & 2689 & 0.1751 \\
\texttt{Winwebsec} & 2303  & 0.1500 \\
\texttt{Renos} & 1567 & 0.1020 \\
\texttt{OnLineGames} & 1511 & 0.0984 \\
\texttt{BHO} & 1412 & 0.0920 \\
\texttt{Startpage} & 1347 & 0.0877 \\
\texttt{Adload} & 1225 & 0.0798 \\
\texttt{VB} & 1110 & 0.0723 \\
\texttt{Vobfus} & 1108  & 0.0721 \\
\texttt{Ceeinject} & 1084 & 0.0706 \\
\midrule
Total & 15,356 & 1.0000 \\ \midrule\midrule
\end{tabular}
}
\end{table}

Next, we describe each of these families. Note that several different classes
of malware are represented, including viruses, worms, and Trojans.

\begin{description}
\item[\texttt{VBinject}\textrm{,}\!] short for ``Visual Basic Injection'', is a
general technique that is applied by malware author to inject malicious program into legitimate 
Windows processes~\cite{VBinject}. This technique is commonly used by malware to evade 
detection by antivirus software and other security measures. Once the malware is injected, 
it can carry out a variety of malicious actions, such as stealing sensitive information, 
downloading additional malware, or taking control of the infected system.

\item[\texttt{Winwebsec}\!] is designed to trick users into purchasing fraudulent security software or 
services by displaying false alerts and warnings about supposed security threats on their computers.
Once installed on a user's computer, \texttt{Winwebsec} will typically display fake warnings claiming 
that the system is infected with viruses, spyware, or other malicious software. These warnings are 
often accompanied by instructions to download and install a security program or pay for a service 
to remove the alleged threats~\cite{Winwebsec}. \texttt{Winwebsec} is often distributed through 
social engineering tactics such as spam emails, malicious websites, and file-sharing networks. 

\item[\texttt{Renos}\!] is similar to Winwebsec, in that it is designed to trick users into purchasing 
fraudulent security software or services~\cite{Renos}. Like other types of fake antivirus malware, 
\texttt{Renos} typically display fake warnings claiming that the system is infected with viruses, 
spyware, or other malicious software, and these warnings are often accompanied by instructions 
to download (and pay for) a supposed anti-virus program. Renos is distributed in the same manner
as Winwebsec. 

\item[\texttt{OnLineGames}\textrm{,}\!] as its name suggests, is a Trojan that mimics an online game,
but is is actuall designed to steal user information. This malware is often distributed through malicious 
websites, peer-to-peer networks, or email attachments. \texttt{OnLineGames} may be particularly dangerous 
because it targets a vulnerable population of online gamers who may be less aware of the risks 
associated with downloading and installing unknown software. Additionally, this type of malware 
can be difficult to detect and remove because it often operates in the background and can evade 
detection by antivirus software~\cite{OnLineGames}.

\item[\texttt{BHO}\textrm{,}\!] which is short for ``Browser Helper Object'', is a type of add-on or 
plugin for web browsers, such as Internet Explorer. Legitimate BHOs provide additional functionality 
or modify the behavior of the browser; however, this \texttt{BHO} malware can be used by 
to perform unwanted actions, such as redirecting web traffic or displaying unwanted ads~\cite{BHO}. 
Because a \texttt{BHO} has deep access to the browser's functionality, it can be difficult to remove once 
installed. In some cases, a malicious \texttt{BHO} may be bundled with legitimate software and installed 
without the user's knowledge or consent.

\item[\texttt{Startpage}\!] is a family is Trojans that modifies a user's web browser settings, such as 
the homepage and search engine, without the user's consent~\cite{Startpage}. Once installed, 
it changes the browser settings to redirect the user's searches to a specific search engine or homepage 
that may contain advertisements or other unwanted content. In some cases, this browser hijacker 
may also install additional unwanted software or collect information about the user's browsing habits.

\item[\texttt{Adload}\!] is an adware program that displays unwanted advertisements that the user cannot 
control as they browse the web~\cite{Adload}. This malware may also collect information about the 
user's browsing habits and use this data to display targeted advertisements. \texttt{Adload} can be difficult 
to remove and may continue to display unwanted advertisements even after the user has attempted to 
uninstall the software. In some cases, it may also install additional malware or compromise the security 
of the victim's computer.

\item[\texttt{VB}\!] is short for ``Visual Basic'', and it is a simple Trojan. It spreads a worm by copying 
itself to removable drives, network shares, and other accessible file systems. Once installed on a victim's 
computer, \texttt{VB} may perform a variety of malicious actions, such as stealing sensitive information, 
logging keystrokes, downloading additional malware, or using the victim's computer to participate in botnets 
or distributed denial-of-service (DDoS) attacks. It is particularly dangerous as it spreads rapidly and may 
infect a large number of computers before it is detected~\cite{VB}. 

\item[\texttt{Vobfus}\!] is a malware family that downloads other malware onto a victim’s computer, including 
\texttt{Beebone}, \texttt{Fareit}, and \texttt{Zbot}. It spreads through infected USB drives, network shares, 
and malicious URLs, and is known for its ability to mutate and evade detection by security software~\cite{Vobfus}. 
\texttt{Vobfus} is dangerous, in part, because it can propagate rapidly and silently, making it difficult to detect 
and contain. It can also disable or bypass security software, making it challenging to remove. 

\item[\texttt{Ceeinject}\!] injects itself into legitimate processes running on a Windows operating system, 
allowing it to execute its malicious code undetected. It is often used in conjunction with other malware, 
such as banking Trojans, to steal sensitive information from victims. This particular threat employs 
obfuscation techniques to conceal its true intentions, making it more difficult for security software to detect its 
malicious activities~\cite{Ceeinject}.
\end{description}

For our feature vectors, we extract a relative byte histogram from each sample. That is, 
given a sample~$S$ in the form of an \texttt{exe} file, we count the
number of times that each byte value~0 through~255 occurs in~$S$, and
then divide each of these counts by the total number of bytes in~$S$. Note that
this implies that our feature vectors are all of length~256. Also, if~$s_i$ is the~$i^{\thth}$
component of the feature vector for the sample~$S$, then~$s_i$ can be interpreted as the 
probability of drawing byte value~$i$, when randomly selecting a byte from~$S$.
These feature vectors are efficient to generate, and require no costly
disassembly or dynamic analysis.

\subsection{Model Training}

Analogous training and testing procedures were used for all learning models considered. 
For the first step, we train each model with labeled data and test the model, which establishes 
a baseline level of performance. We use accuracy as our measure of performance.
 
After the initial training and testing, data is inserted into the low-order~$n$ bits of the weights, 
which, on average, changes about half of the bit values. For each~$n$, the performance of the 
model is re-evaluated using the same data and accuracy metric as previously used,
which allows for a direct comparison of the results for each~$n$. We then graph these accuracy
results as a function of~$n$.

\section{Steganographic Capacity Experiments}\label{chap:4}

In this section, we consider the steganographic capacity of each of the models discussed in
Section~\ref{sect:learningModels}. To measure the steganographic capacity, 
we embed information in the low-order~$n$ bits of the model weights. 
For each model, we consider the following three cases.
\begin{enumerate}
\item Only the output layer weights are modified
\item Only the weights of the hidden layer (or layers) are modified
\item All of the model weights are modified
\end{enumerate}
In each case, we graph the model accuracy as a function of~$n$. Also, 
we discuss the total capacity, that is, the total number of model bits that 
are available for this form of information hiding. In each case, the information that 
we hide is extracted from the book \textit{Alice's Adventures in Wonderland}~\cite{Alice}.

\subsection{MLP}

The {\tt MLPClassifier()} from the \texttt{sklearn.neural\_network} module was used to 
train and test our MLP model. The hyperparameters tested are listed in Table~\ref{tab:3}, 
with the selected values in boldface. Note that a model with two hidden layers, 
with~128 and~10 neurons, respectively, was best. Also, the logistic function was 
selected as our activation function, and so on.   

\begin{table}[!htb]
\caption{MLP model hyperparameters tested}\label{tab:3}
\centering
\adjustbox{scale=0.85}{
\begin{tabular}{c|c}\midrule\midrule
Hyperparameter & Values tested\\ \midrule
\texttt{hidden\_layer\_sizes} & (64, 10), (96, 10), \textbf{(128, 10)} \\
\texttt{activation} & identity, \texttt{\textbf{logistic}}  \\
\texttt{alpha} & 0.0001, \textbf{0.05} \\
\texttt{random\_state} & 30, \textbf{40}, 50 \\ 
\texttt{solver} & \texttt{\textbf{adam}} \\
\texttt{learning\_rate\_init} &  \textbf{0.00001} \\
\texttt{max\_iter} &  \textbf{10000} \\\midrule\midrule
\end{tabular}
}
\end{table}

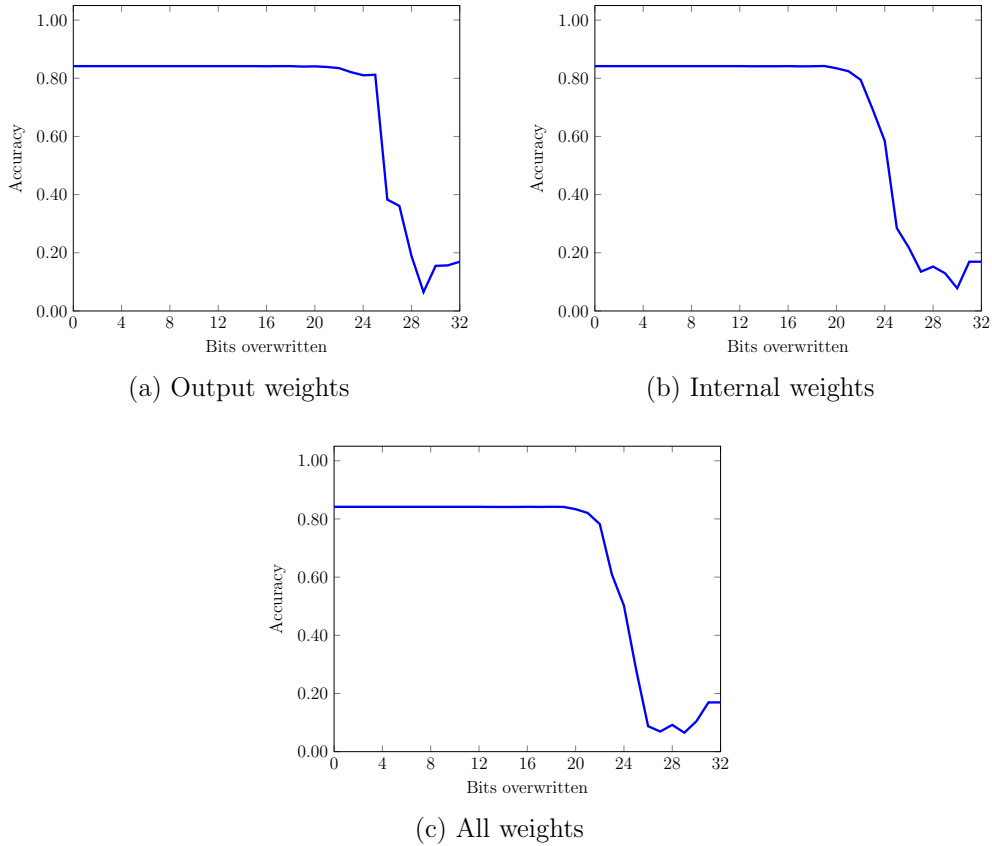
\begin{figure}[!htb]
\centering
\begin{tabular}{cc}
\adjustbox{scale=0.925}{
\begin{tikzpicture}[scale=0.6]
\begin{axis}[ 
		   width=0.7\textwidth,
		   height=0.575\textwidth,
	 	   x tick label style={
   		 	/pgf/number format/.cd,
			/pgf/number format/1000 sep={},
   			fixed,
   			fixed zerofill,
    			precision=0
		   },
	 	   y tick label style={
    		 	/pgf/number format/.cd,
   			fixed,
   			fixed zerofill,
    			precision=2
		    },
                    xmin=0,xmax=32,
                    ymin=0.00,ymax=1.05,
                    legend pos=south west,
                    xtick={0,4,8,12,16,20,24,28,32},
                    ytick={0.00,0.20,0.40,0.60,0.80,1.00},
                    xlabel={Bits overwritten},
                    ylabel={Accuracy}] 
\addplot[color=blue,ultra thick,mark=none] coordinates { 
(0, 0.8416768416768416)
(1, 0.8416768416768416)
(2, 0.8416768416768416)
(3, 0.8416768416768416)
(4, 0.8416768416768416)
(5, 0.8416768416768416)
(6, 0.8416768416768416)
(7, 0.8416768416768416)
(8, 0.8416768416768416)
(9, 0.8416768416768416)
(10, 0.8416768416768416)
(11, 0.8416768416768416)
(12, 0.8416768416768416)
(13, 0.8416768416768416)
(14, 0.8416768416768416)
(15, 0.8416768416768416)
(16, 0.8412698412698413)
(17, 0.8416768416768416)
(18, 0.8416768416768416)
(19, 0.8400488400488401)
(20, 0.8408628408628409)
(21, 0.8388278388278388)
(22, 0.8347578347578347)
(23, 0.820919820919821)
(24, 0.8103378103378104)
(25, 0.811965811965812)
(26, 0.382987382987383)
(27, 0.361009361009361)
(28, 0.18925518925518925)
(29, 0.06512006512006512)
(30, 0.15506715506715507)
(31, 0.15669515669515668)
(32, 0.1693121693121693)};
\end{axis}
\end{tikzpicture}
}
&
\adjustbox{scale=0.925}{
\begin{tikzpicture}[scale=0.6]
\begin{axis}[ 
		   width=0.7\textwidth,
		   height=0.575\textwidth,
	 	   x tick label style={
   		 	/pgf/number format/.cd,
			/pgf/number format/1000 sep={},
   			fixed,
   			fixed zerofill,
    			precision=0
		   },
	 	   y tick label style={
    		 	/pgf/number format/.cd,
   			fixed,
   			fixed zerofill,
    			precision=2
		    },
                    xmin=0,xmax=32,
                    ymin=0.00,ymax=1.05,
                    legend pos=south west,
                    xtick={0,4,8,12,16,20,24,28,32},
                    ytick={0.00,0.20,0.40,0.60,0.80,1.00},
                    xlabel={Bits overwritten},
                    ylabel={Accuracy}] 
\addplot[color=blue,ultra thick,mark=none] coordinates { 
(0, 0.8416768416768416)
(1, 0.8416768416768416)
(2, 0.8416768416768416)
(3, 0.8416768416768416)
(4, 0.8416768416768416)
(5, 0.8416768416768416)
(6, 0.8416768416768416)
(7, 0.8416768416768416)
(8, 0.8416768416768416)
(9, 0.8416768416768416)
(10, 0.8416768416768416)
(11, 0.8416768416768416)
(12, 0.8416768416768416)
(13, 0.8412698412698413)
(14, 0.8412698412698413)
(15, 0.8412698412698413)
(16, 0.8416768416768416)
(17, 0.8408628408628409)
(18, 0.8412698412698413)
(19, 0.8420838420838421)
(20, 0.8343508343508343)
(21, 0.8241758241758241)
(22, 0.7948717948717948)
(23, 0.6931216931216931)
(24, 0.5844525844525844)
(25, 0.2849002849002849)
(26, 0.21733821733821734)
(27, 0.13512413512413513)
(28, 0.15262515262515264)
(29, 0.12942612942612944)
(30, 0.07814407814407814)
(31, 0.1693121693121693)
(32, 0.1693121693121693)};
\end{axis}
\end{tikzpicture}
}
\\
\adjustbox{scale=0.85}{(a) Output weights}
&
\adjustbox{scale=0.85}{(b) Internal weights}
\\ \\
\multicolumn{2}{c}{\adjustbox{scale=0.925}{
\begin{tikzpicture}[scale=0.6]
\begin{axis}[ 
		   width=0.7\textwidth,
		   height=0.575\textwidth,
	 	   x tick label style={
   		 	/pgf/number format/.cd,
			/pgf/number format/1000 sep={},
   			fixed,
   			fixed zerofill,
    			precision=0
		   },
	 	   y tick label style={
    		 	/pgf/number format/.cd,
   			fixed,
   			fixed zerofill,
    			precision=2
		    },
                    xmin=0,xmax=32,
                    ymin=0.00,ymax=1.05,
                    legend pos=south west,
                    xtick={0,4,8,12,16,20,24,28,32},
                    ytick={0.00,0.20,0.40,0.60,0.80,1.00},
                    xlabel={Bits overwritten},
                    ylabel={Accuracy}] 
\addplot[color=blue,ultra thick,mark=none] coordinates { 
(0, 0.8416768416768416)
(1, 0.8416768416768416)
(2, 0.8416768416768416)
(3, 0.8416768416768416)
(4, 0.8416768416768416)
(5, 0.8416768416768416)
(6, 0.8416768416768416)
(7, 0.8416768416768416)
(8, 0.8416768416768416)
(9, 0.8416768416768416)
(10, 0.8416768416768416)
(11, 0.8416768416768416)
(12, 0.8416768416768416)
(13, 0.8412698412698413)
(14, 0.8412698412698413)
(15, 0.8412698412698413)
(16, 0.8416768416768416)
(17, 0.8412698412698413)
(18, 0.8416768416768416)
(19, 0.8412698412698413)
(20, 0.8335368335368335)
(21, 0.8205128205128205)
(22, 0.7822547822547823)
(23, 0.6092796092796092)
(24, 0.5018315018315018)
(25, 0.2836792836792837)
(26, 0.08750508750508751)
(27, 0.06919006919006919)
(28, 0.09198209198209198)
(29, 0.06512006512006512)
(30, 0.1045991045991046)
(31, 0.1693121693121693)
(32, 0.1693121693121693)};
\end{axis}
\end{tikzpicture}
}}
\\
\multicolumn{2}{c}{\adjustbox{scale=0.85}{(c) All weights}}
\end{tabular}
\caption{MLP model performance with low-order bits of weights overwritten}\label{fig:MLP_Plot}
\end{figure}

The results obtained when hiding information in the low order bits of the output layer weights 
of our trained MLP model are summarized in Figure~\ref{fig:MLP_Plot}(a). We observe that
the original accuracy for the model is approximately~0.8417, and the performance of the model 
exceeds~0.8119, until the low-order~26 bits of the output weights
are overwritten, which causes the accuracy to drop dramatically 
to~0.3830. Overwriting more bits causes the accuracy to fluctuate, but it remains very low.

In Figures~\ref{fig:MLP_Plot}(b) and~(c) we give the results when information is hidden 
in the hidden layer weights, and when information is hidden in all of the weights of
our trained MLP model, respectively. The results in these two cases are analogous
to the results for the output layer weights, although in both of these latter
cases, only~21 bits can be overwritten before the accuracy drops below~0.80.

There are~100 weights in the output layer, and~34,048 weights in the hidden layer, 
which makes the total number of weights~34,148 in this particular MLP model. 
As shown in the results in Figure~\ref{fig:MLP_Plot}(a), we can overwrite the low-order-25 bits 
of each weights in the output layer with minimal loss of accuracy, which
gives the model a steganographic capacity of 2.44KB,\footnote{Note that we follow the 
standard convention in computing whereby~1KB represents~$2^{10}$ bytes, 
while~1MB is~$2^{20}$ bytes, and~1GB is~$2^{30}$ bytes.} just in the output layer. 
The results in Figure~\ref{fig:MLP_Plot}(b) show that inserting information into the 
low-order-21 bits weights in the internal layers does not have a major
negative impact on the model accuracy, which gives the a steganographic capacity 
of slightly more than~698KB. With all weights in the model considered, 
as shown in Figure~\ref{fig:MLP_Plot}(c), again the low-order-21 bits are available for information hiding, 
which give the MLP model a steganographic capacity that is slightly in excess of~700KB. 

\subsection{CNN}

Our CNN model was implemented using {\tt torch} in PyTorch, which provides support for 
tensor computation, deep neural networks, and many other useful machine learning utilities. 
The model architecture selected consists of two convolutional layers, each utilizing 
ReLU activation functions, with one and six input channels, as well as six 
and~12 output channels, respectively. Following the convolutional layers, 
there are two fully connected linear layers, again with ReLU activation functions. 
The input sizes of these fully connected layers are~$12\times 256$ and~512, respectively. 
The final layer of the model is a fully connected output layer with an input size of~100 
and an output size of~10, utilizing a linear activation function. 

The hyperparameters tested (via grid search)
are listed in detail in Table~\ref{tab:4}, with the selected values in boldface.
Since this model has a large number of hyperparameters and training is relatively
costly, only two of the hyperparameter values are varied

\begin{table}[!htb]
\caption{CNN model hyperparameters tested}\label{tab:4}
\centering
\adjustbox{scale=0.85}{
\begin{tabular}{c|c}\midrule\midrule
Hyperparameter & Values tested \\ \midrule
\texttt{pad-size} & \textbf{256} \\
\texttt{batch size} & 128, \textbf{64} \\
\texttt{max-epoch} & \textbf{20} \\
\texttt{lr} & 0.0005, \textbf{0.00005} \\
\texttt{momentum} &  \textbf{0.9} \\
\texttt{hidden-size} & \textbf{512} \\
\texttt{output} &  \textbf{10} \\
\texttt{bptt} &  \textbf{256} \\
\texttt{ntoken} &  \textbf{256} \\
\texttt{d\_model} &  \textbf{128} \\
\texttt{d\_hid} &  \textbf{128} \\
\texttt{nlayers} &  \textbf{2} \\
\texttt{nhead} &  \textbf{1} \\
\texttt{dropout} &  \textbf{0.5} \\ \midrule\midrule
\end{tabular}
}
\end{table}

As with the previous models, process of analyzing the impact of hiding information in the output layer weights 
on the accuracy of the CNN model was carried out systematically. The model was initially trained with the 
preprocessed malware family dataset, and its accuracy was evaluated on the testing data. The accuracy 
was found to be~0.7354 in this unmodified case, which serves as the baseline for subsequent analysis.

\begin{figure}[!htb]
\centering
\begin{tabular}{cc}
\adjustbox{scale=0.925}{
\begin{tikzpicture}[scale=0.6]
\begin{axis}[ 
		   width=0.7\textwidth,
		   height=0.575\textwidth,
	 	   x tick label style={
   		 	/pgf/number format/.cd,
			/pgf/number format/1000 sep={},
   			fixed,
   			fixed zerofill,
    			precision=0
		   },
	 	   y tick label style={
    		 	/pgf/number format/.cd,
   			fixed,
   			fixed zerofill,
    			precision=2
		    },
                    xmin=0,xmax=32,
                    ymin=0.00,ymax=1.05,
                    legend pos=south west,
                    xtick={0,4,8,12,16,20,24,28,32},
                    ytick={0.00,0.20,0.40,0.60,0.80,1.00},
                    xlabel={Bits overwritten},
                    ylabel={Accuracy}] 
\addplot[color=blue,ultra thick,mark=none] coordinates { 
(0, 0.7501017501017501)
(1, 0.7501017501017501)
(2, 0.7501017501017501)
(3, 0.7501017501017501)
(4, 0.7501017501017501)
(5, 0.7501017501017501)
(6, 0.7501017501017501)
(7, 0.7501017501017501)
(8, 0.7501017501017501)
(9, 0.7501017501017501)
(10, 0.7501017501017501)
(11, 0.7501017501017501)
(12, 0.7501017501017501)
(13, 0.7501017501017501)
(14, 0.7501017501017501)
(15, 0.7496947496947497)
(16, 0.7505087505087505)
(17, 0.7484737484737485)
(18, 0.7488807488807488)
(19, 0.7484737484737485)
(20, 0.7533577533577533)
(21, 0.7468457468457469)
(22, 0.706959706959707)
(23, 0.7232397232397232)
(24, 0.5575905575905576)
(25, 0.5091575091575091)
(26, 0.5335775335775336)
(27, 0.25071225071225073)
(28, 0.20146520146520147)
(29, 0.13797313797313798)
(30, 0.20512820512820512)
(31, 0.08628408628408628)
(32, 0.15466015466015465)};
\end{axis}
\end{tikzpicture}
}
&
\adjustbox{scale=0.925}{
\begin{tikzpicture}[scale=0.6]
\begin{axis}[ 
		   width=0.7\textwidth,
		   height=0.575\textwidth,
	 	   x tick label style={
   		 	/pgf/number format/.cd,
			/pgf/number format/1000 sep={},
   			fixed,
   			fixed zerofill,
    			precision=0
		   },
	 	   y tick label style={
    		 	/pgf/number format/.cd,
   			fixed,
   			fixed zerofill,
    			precision=2
		    },
                    xmin=0,xmax=32,
                    ymin=0.00,ymax=1.05,
                    legend pos=south west,
                    xtick={0,4,8,12,16,20,24,28,32},
                    ytick={0.00,0.20,0.40,0.60,0.80,1.00},
                    xlabel={Bits overwritten},
                    ylabel={Accuracy}] 
\addplot[color=blue,ultra thick,mark=none] coordinates { 
(0, 0.7501017501017501)
(1, 0.7501017501017501)
(2, 0.7501017501017501)
(3, 0.7501017501017501)
(4, 0.7501017501017501)
(5, 0.7501017501017501)
(6, 0.7501017501017501)
(7, 0.7501017501017501)
(8, 0.7501017501017501)
(9, 0.7501017501017501)
(10, 0.7501017501017501)
(11, 0.7501017501017501)
(12, 0.7501017501017501)
(13, 0.7496947496947497)
(14, 0.7501017501017501)
(15, 0.7505087505087505)
(16, 0.7501017501017501)
(17, 0.7484737484737485)
(18, 0.7492877492877493)
(19, 0.7496947496947497)
(20, 0.7492877492877493)
(21, 0.7639397639397639)
(22, 0.7505087505087505)
(23, 0.7191697191697192)
(24, 0.2498982498982499)
(25, 0.24094424094424094)
(26, 0.24175824175824176)
(27, 0.08017908017908018)
(28, 0.07326007326007326)
(29, 0.07163207163207164)
(30, 0.1225071225071225)
(31, 0.1693121693121693)
(32, 0.1693121693121693)};
\end{axis}
\end{tikzpicture}
}
\\
\adjustbox{scale=0.85}{(a) Output weights}
&
\adjustbox{scale=0.85}{(b) Internal weights}
\\ \\
\multicolumn{2}{c}{\adjustbox{scale=0.925}{
\begin{tikzpicture}[scale=0.6]
\begin{axis}[ 
		   width=0.7\textwidth,
		   height=0.575\textwidth,
	 	   x tick label style={
   		 	/pgf/number format/.cd,
			/pgf/number format/1000 sep={},
   			fixed,
   			fixed zerofill,
    			precision=0
		   },
	 	   y tick label style={
    		 	/pgf/number format/.cd,
   			fixed,
   			fixed zerofill,
    			precision=2
		    },
                    xmin=0,xmax=32,
                    ymin=0.00,ymax=1.05,
                    legend pos=south west,
                    xtick={0,4,8,12,16,20,24,28,32},
                    ytick={0.00,0.20,0.40,0.60,0.80,1.00},
                    xlabel={Bits overwritten},
                    ylabel={Accuracy}] 
\addplot[color=blue,ultra thick,mark=none] coordinates { 
(0, 0.7501017501017501)
(1, 0.7501017501017501)
(2, 0.7501017501017501)
(3, 0.7501017501017501)
(4, 0.7501017501017501)
(5, 0.7501017501017501)
(6, 0.7501017501017501)
(7, 0.7501017501017501)
(8, 0.7501017501017501)
(9, 0.7501017501017501)
(10, 0.7501017501017501)
(11, 0.7501017501017501)
(12, 0.7501017501017501)
(13, 0.7496947496947497)
(14, 0.7501017501017501)
(15, 0.7505087505087505)
(16, 0.7501017501017501)
(17, 0.7484737484737485)
(18, 0.7492877492877493)
(19, 0.7496947496947497)
(20, 0.7492877492877493)
(21, 0.7639397639397639)
(22, 0.7505087505087505)
(23, 0.7191697191697192)
(24, 0.2498982498982499)
(25, 0.24094424094424094)
(26, 0.24175824175824176)
(27, 0.08017908017908018)
(28, 0.07326007326007326)
(29, 0.07163207163207164)
(30, 0.1225071225071225)
(31, 0.1693121693121693)
(32, 0.1693121693121693)};
\end{axis}
\end{tikzpicture}
}}
\\
\multicolumn{2}{c}{\adjustbox{scale=0.85}{(c) All weights}}
\end{tabular}
\caption{CNN model performance with low-order bits of weights overwritten}\label{fig:CNN_Plot}
\end{figure}
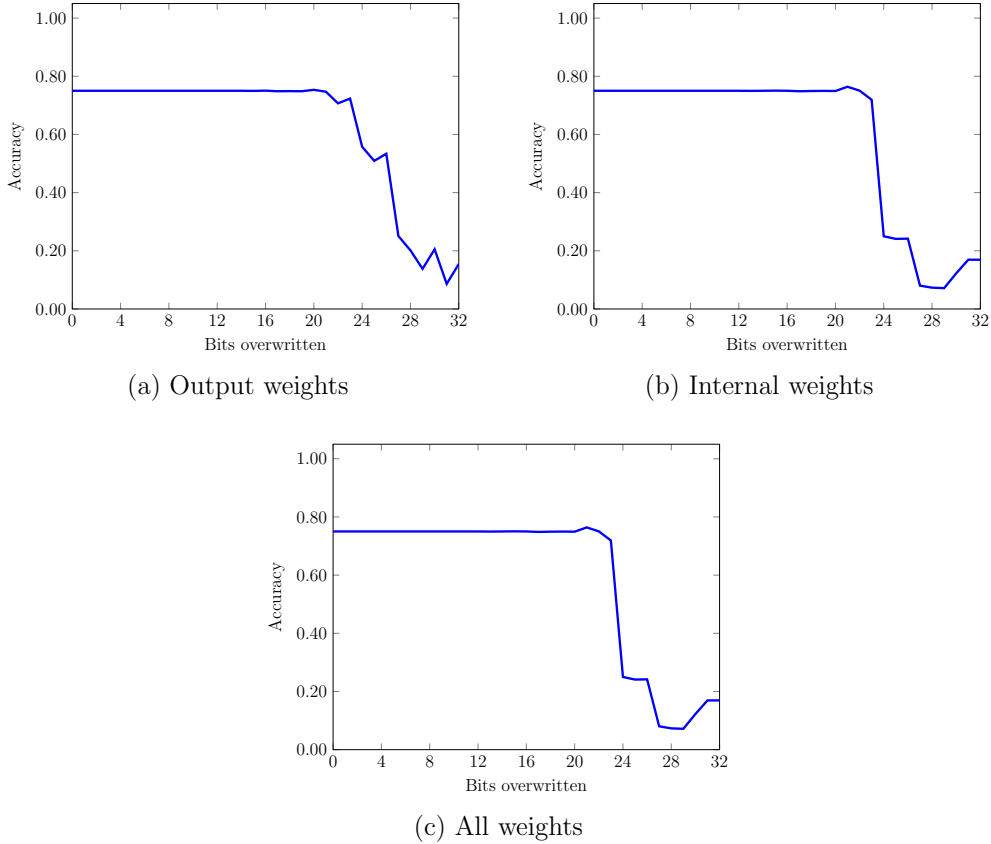

Next, the output layer weights were systematically overwritten with data, starting from the low-order
bits and increasing towards the high-order bits. A total of 32 bits are present in each weight, 
and the resulting accuracy was recorded after the~$n$ low-order bits had been overwritten,
for each~$n\in\{0,1,2,\ldots,32\}$. Figure~\ref{fig:CNN_Plot}(a) summarizes the accuracies
obtained for the model in each case.

We observe that overwriting the low-order~21 bits of the output layer weights does 
not have any significant effect on the accuracy. However, when the~$22^{\ndnd}$ bit is 
overwritten, the accuracy drops from~0.7468 to~0.7070,
and a large drop to~0.5576 occurs when the low-order~24 bits are 
overwritten. Finally, another large drop is accuracy is observed when the~27 low-order
bits are overwritten, resulting in an accuracy of only~0.2507, and when~29
low-order bits are overwritten, the accuracy is comparable
to guessing the labels at random.

In Figures~\ref{fig:CNN_Plot}(b) and~(c) we give the results when information is hidden 
in the hidden layer weights, and when information is hidden in all of the weights
of our trained CNN model, respectively. These results are analogous to the
output layer case, but with~22 low-order bits available for information hiding
in both, and a sharper drop in accuracy from that point.

In this particular CNN model, there are~1000 weights in the output layer, 
and~1,624,142 weights in the internal layer, and hence the total number of weights is~1,625,142.
As shown in the results in Figure~\ref{fig:CNN_Plot}(a), we can change the low-order-21 bits of each 
weight in the output layer without significantly affecting the model performance, which
gives the model a steganographic capacity of slightly more than~20.5KB, just in terms of the output layer. 
The results in Figure~\ref{fig:CNN_Plot}(b) show that inserting information into the low-order-22 bits 
of the weights in the internal layers does not have a negative impact on the model accuracy, 
which gives the model a steganographic capacity of about~34.0MB in terms of the internal weights. 
With all weights of the model considered, as shown in Figure~\ref{fig:CNN_Plot}(c), 
the low-order~22 bits are again available for information hiding, which give the MLP model 
a total steganographic capacity of about~34.1MB. 

\subsection{Transformer Model}

Our Transformer model was implemented using the PyTorch modules 
including {\tt TransformerEncoder}, {\tt TransformerEncoderLayer}, 
{\tt TransformerDecoderLayer}, {\tt TransformerDecoder}, and {\tt LayerNorm}. 
The model consists of an embedding layer, a positional encoding layer, a Transformer encoder layer, 
a Transformer decoder layer, and two linear layers. The input first passes through an embedding layer, 
which maps each token in the input sequence to a vector in a high-dimensional space. 
Then, a positional encoding layer is applied to the embedded input sequence to add 
positional information to the embeddings.

The Transformer encoder layer serves to encode the input sequence and create 
a representation of it in a high-dimensional space. The encoder layer is composed of 
a self-attention mechanism and a feedforward neural network layer. The resulting vectors 
are then passed through a feedforward neural network layer.

Similarly, the Transformer decoder layer takes the encoded input sequence and generates 
a prediction for each output token. The decoder layer is also composed of self-attention 
and feedforward neural network layers. However, it additionally received inputs from the 
encoder layer through a multi-head attention mechanism.

Finally, the output of the Transformer decoder layer is passed through two linear layers, 
where the first layer maps the output to a lower-dimensional space, and the second layer 
maps this lower-dimensional representation to the output classes. The model also 
employs layer normalization and dropout for regularization.

The hyperparameters tested via a grid search are listed in Table~\ref{tab:5},
with the values selected in boldface. Since there are a large number of
hyperparameters in this model, seven of the hyperparameters in Table~\ref{tab:5}
are fixed values. Our trained transformer model achieved perfect accuracy on the test dataset. 

\begin{table}[!htb]
\caption{Transformer model hyperparameters tested}\label{tab:5}
\centering
\adjustbox{scale=0.85}{
\begin{tabular}{c|c}\midrule\midrule
Hyperparameter & Values tested \\ \midrule
\texttt{pad-size} & \textbf{256} \\
\texttt{batch size} & 128, \textbf{64} \\
\texttt{max-epoch} & \textbf{20} \\
\texttt{lr} & 0.0005, \textbf{0.00005} \\
\texttt{momentum} &  \textbf{0.9} \\
\texttt{hidden-size} & 256, \textbf{512} \\
\texttt{output} &  \textbf{10} \\
\texttt{bptt} &  \textbf{256} \\
\texttt{ntoken} &  \textbf{256} \\
\texttt{d\_model} &  \textbf{128}, 256, 512 \\
\texttt{d\_hid} &  \textbf{128}, 1024 \\
\texttt{nlayers} &  \textbf{2}, 12 \\
\texttt{nhead} &  \textbf{1}, 8 \\
\texttt{dropout} &  \textbf{0.5} \\ \midrule\midrule
\end{tabular}
}
\end{table}

The weights of the output layer were manipulated to explore the effect of overwriting
the low-order bits. The resulting accuracies---as a function of the number of bits 
overwritten---can be seen in Figure~\ref{fig:TF_Plot}(a).
We observe that up to~26 low-order bits can be overwritten with
no adverse effect on the model accuracy. A drop in accuracy from~1.00 to~0.9833
occurs when the low-order-27 bits are overwritten. When the low-order~28 bits 
of the output layer weights are overwritten, the accuracy of the model drops to~0.8307,
and the accuracy plummets thereafter. 


\begin{figure}[!htb]
\centering
\begin{tabular}{cc}
\adjustbox{scale=0.925}{
\begin{tikzpicture}[scale=0.6]
\begin{axis}[ 
		   width=0.7\textwidth,
		   height=0.575\textwidth,
	 	   x tick label style={
   		 	/pgf/number format/.cd,
			/pgf/number format/1000 sep={},
   			fixed,
   			fixed zerofill,
    			precision=0
		   },
	 	   y tick label style={
    		 	/pgf/number format/.cd,
   			fixed,
   			fixed zerofill,
    			precision=2
		    },
                    xmin=0,xmax=32,
                    ymin=0.00,ymax=1.05,
                    legend pos=south west,
                    xtick={0,4,8,12,16,20,24,28,32},
                    ytick={0.00,0.20,0.40,0.60,0.80,1.00},
                    xlabel={Bits overwritten},
                    ylabel={Accuracy}] 
\addplot[color=blue,ultra thick,mark=none] coordinates { 
(0, 1.0)
(1, 1.0)
(2, 1.0)
(3, 1.0)
(4, 1.0)
(5, 1.0)
(6, 1.0)
(7, 1.0)
(8, 1.0)
(9, 1.0)
(10, 1.0)
(11, 1.0)
(12, 1.0)
(13, 1.0)
(14, 1.0)
(15, 1.0)
(16, 1.0)
(17, 1.0)
(18, 1.0)
(19, 1.0)
(20, 1.0)
(21, 1.0)
(22, 1.0)
(23, 1.0)
(24, 1.0)
(25, 1.0)
(26, 1.0)
(27, 0.9833129833129833)
(28, 0.8306878306878307)
(29, 0.6418396418396418)
(30, 0.38217338217338215)
(31, 0.06512006512006512)
(32, 0.15466015466015465)};
\end{axis}
\end{tikzpicture}
}
&
\adjustbox{scale=0.925}{
\begin{tikzpicture}[scale=0.6]
\begin{axis}[ 
		   width=0.7\textwidth,
		   height=0.575\textwidth,
	 	   x tick label style={
   		 	/pgf/number format/.cd,
			/pgf/number format/1000 sep={},
   			fixed,
   			fixed zerofill,
    			precision=0
		   },
	 	   y tick label style={
    		 	/pgf/number format/.cd,
   			fixed,
   			fixed zerofill,
    			precision=2
		    },
                    xmin=0,xmax=32,
                    ymin=0.00,ymax=1.05,
                    legend pos=south west,
                    xtick={0,4,8,12,16,20,24,28,32},
                    ytick={0.00,0.20,0.40,0.60,0.80,1.00},
                    xlabel={Bits overwritten},
                    ylabel={Accuracy}] 
\addplot[color=blue,ultra thick,mark=none] coordinates { 
(0, 1.0)
(1, 1.0)
(2, 1.0)
(3, 1.0)
(4, 1.0)
(5, 1.0)
(6, 1.0)
(7, 1.0)
(8, 1.0)
(9, 1.0)
(10, 1.0)
(11, 1.0)
(12, 1.0)
(13, 1.0)
(14, 1.0)
(15, 1.0)
(16, 1.0)
(17, 1.0)
(18, 1.0)
(19, 1.0)
(20, 1.0)
(21, 1.0)
(22, 1.0)
(23, 1.0)
(24, 1.0)
(25, 0.6251526251526252)
(26, 0.11518111518111518)
(27, 0.15913715913715915)
(28, 0.09727309727309727)
(29, 0.1693121693121693)
(30, 0.1693121693121693)
(31, 0.1693121693121693)
(32, 0.1693121693121693)};
\end{axis}
\end{tikzpicture}
}
\\
\adjustbox{scale=0.85}{(a) Output weights}
&
\adjustbox{scale=0.85}{(b) Internal weights}
\\ \\
\multicolumn{2}{c}{\adjustbox{scale=0.925}{
\begin{tikzpicture}[scale=0.6]
\begin{axis}[ 
		   width=0.7\textwidth,
		   height=0.575\textwidth,
	 	   x tick label style={
   		 	/pgf/number format/.cd,
			/pgf/number format/1000 sep={},
   			fixed,
   			fixed zerofill,
    			precision=0
		   },
	 	   y tick label style={
    		 	/pgf/number format/.cd,
   			fixed,
   			fixed zerofill,
    			precision=2
		    },
                    xmin=0,xmax=32,
                    ymin=0.00,ymax=1.05,
                    legend pos=south west,
                    xtick={0,4,8,12,16,20,24,28,32},
                    ytick={0.00,0.20,0.40,0.60,0.80,1.00},
                    xlabel={Bits overwritten},
                    ylabel={Accuracy}] 
\addplot[color=blue,ultra thick,mark=none] coordinates { 
(0, 1.0)
(1, 1.0)
(2, 1.0)
(3, 1.0)
(4, 1.0)
(5, 1.0)
(6, 1.0)
(7, 1.0)
(8, 1.0)
(9, 1.0)
(10, 1.0)
(11, 1.0)
(12, 1.0)
(13, 1.0)
(14, 1.0)
(15, 1.0)
(16, 1.0)
(17, 1.0)
(18, 1.0)
(19, 1.0)
(20, 1.0)
(21, 1.0)
(22, 1.0)
(23, 1.0)
(24, 1.0)
(25, 0.31094831094831094)
(26, 0.09686609686609686)
(27, 0.13838013838013838)
(28, 0.1339031339031339)
(29, 0.1693121693121693)
(30, 0.1693121693121693)
(31, 0.1693121693121693)
(32, 0.1693121693121693)};
\end{axis}
\end{tikzpicture}
}}
\\
\multicolumn{2}{c}{\adjustbox{scale=0.85}{(c) All weights}}
\end{tabular}
\caption{Transformer model performance with low-order bits of weights overwritten}\label{fig:TF_Plot}
\end{figure}

In Figures~\ref{fig:TF_Plot}(b) and~(c) we give the results when information is hidden 
in the hidden layer weights, and when information is hidden in all of the weights, of
our trained Transformer model, respectively. In both of these cases, 
we are free to hide information in the low-order~24 bit positions with
no negative effect on the model, but when we use the low-order~25 bits,
model accuracy is severely affected.

In the transformer model, the number of weights in output layer and internal layers 
are~1280 and~175,681,024, respectively, giving a total of~175,682,304 weights. 
For the output layer weights, as shown in Figure~\ref{fig:TF_Plot}(a), we 
can overwrite the low-order~26 bits with minimal loss in accuracy, 
giving a steganographic capacity of~32.5KB, just in terms of the output layer. 
Considering either the internal weights or all weights, we can hide information
in the low~24 bits without any ill effect on the model, giving us
a steganographic capacity in excess of~3.92GB for both cases.

\section{Conclusion}\label{chap:5}

The primary goal of this research was to determine a reasonable
lower bound the stenographic capacity of various learning models. 
Specifically, we tested
Multilayer Perceptron (MLP), 
Convolutional Neural Network (CNN), 
and Transformer models, which were each trained on a dataset of more 
than~15,000 malware executables from~10 families, 
with more than~1000 samples for each family. 

All of the trained deep learning models underwent the same testing procedure:
We first determined the accuracy of each model on the test set, then we embedded information
in the~$n$ low-order bits of the weights, recomputing the classification accuracy for each~$n$. 
We experimented with just the output layer weights, just the hidden layer weights, and all of the weights.
The results were consistent across all models, in the sense that at least~20 bits per 
weight can be used to hide information, with minimal effect on the accuracy. In addition, 
at some point shortly beyond~20 bits, model accuracy deteriorates dramatically. These
results hold whether considering the output layer weights, the hidden layer weights,
or all weights.

Our experimental results show that the steganographic capacity of the deep learning models 
we tested is surprisingly high. This is potentially a significant security issue, 
since such models are ubiquitous, and hence it
is to be expected that attackers will try to take advantage of them. Embedding, say, malware
in a learning model offers an attack vector that is practical, and could be highly effective in practice.

It would be wise to reduce the steganographic capacity of learning models. Our results indicate
that~32-bit weight do not yield a significant improvement in accuracy over what could be achieved
with, say, 16-bit weights. With additional work, for specific models, it should be feasible to use
even smaller weights---this would be an interesting and potentially valuable area for additional 
research.


Further research into other popular deep learning models would also be worthwhile. In particular,
it would be interesting to determine the steganographic capacity of pre-trained models,
such as VGG-19 and any of the popular ResNet models 
(e.g., ResNet18, ResNet34, ResNet50, and ResNet101). Further, it would seem that
the steganographic capacity of pre-trained models could be greatly
reduced, and the creation of ``thin'' pre-trained models would be of value. 
It would also be interesting to determine whether more challenging classification
problems tend to affect the steganographic capacity of inherently ``fat'' models. 
Intuitively, more challenging problems should require more learning to be embedded in
the weights, and hence the steganographic capacity might be somewhat lower.

Another area for further investigation
would be to combine some aspects of the steganographic capacity
work presented in this paper with the work in~\cite{EvilModel}, where
information is hidden in weights that are (essentially) unused by the model.
By combining both of these approaches, we could obtain a larger 
steganographic capacity of learning models, 
with the goal of obtaining a reasonably tight upper bound.

\bibliographystyle{plain}
\bibliography{references.bib}

\begin{thebibliography}{10}

\bibitem{Agarwal_2013}
Monika Agarwal.
\newblock Text stegeganographic approaches: A comparison.
\newblock {\em International Journal of Network Security \&\ Its Applications},
  5(1):91--106, 2013.

\bibitem{ML_Stat}
James Anthony.
\newblock 60 notable machine learning statistics: 2021/2022 market share \&\
  data analysis.
\newblock \url{https://financesonline.com/machine-learning-statistics/}, 2022.

\bibitem{Alice}
Lewis Carroll.
\newblock {\em Alice's Adventures in Wonderland}.
\newblock Macmillan, 1865.
\newblock \url{https://www.gutenberg.org/ebooks/11}.

\bibitem{Clelland_1999}
Catherine~Taylor Clelland, Viviana Risca, and Carter Bancroft.
\newblock Hiding messages in {DNA} microdots.
\newblock {\em Nature}, 399(6736):533--534, 1999.

\bibitem{Whitfield_2022}
Whitfield Diffie and Martin~E. Hellman.
\newblock New directions in cryptography.
\newblock {\em IEEE Transactions on Information Theory}, 22(6):644--654, 1976.

\bibitem{Duggal_2022}
Nikita Duggal.
\newblock Top 10 machine learning applications and examples in 2022.
\newblock
  \url{https://www.simplilearn.com/tutorials/machine-learning-tutorial/machine-learning-applications},
  2022.

\bibitem{Fiscutean_2021}
Andrada Fiscutean.
\newblock Steganography explained and how to protect against it.
\newblock
  \url{https://www.csoonline.com/article/3632146/steganography-explained-and-how-to-protect-against-it.html},
  2021.

\bibitem{John_2017}
John.
\newblock Word of the day: Steganography.
\newblock \url{https://www.secalliance.com/blog/word-day-steganography}, 2017.

\bibitem{VirusShare}
Samuel Kim.
\newblock Pe header analysis for malware detection.
\newblock Master's thesis, San Jose State University, 2018.

\bibitem{Mohammed_2021}
Mohammed~Abdul Majeed, Rossilawati Sulaiman, Zarina Shukur, and Mohammad~Kamrul
  Hasan.
\newblock A review on text steganography techniques.
\newblock {\em Mathematics}, 9(21):2829, 2021.

\bibitem{VB}
{Microsoft}.
\newblock {Trojan:Win32/VB}.
\newblock
  \url{https://www.microsoft.com/en-us/wdsi/threats/malware-encyclopedia-description?Name=Trojan:Win32/VB},
  2007.

\bibitem{Ceeinject}
{Microsoft}.
\newblock {VirTool:Win32/CeeInject}.
\newblock
  \url{https://www.microsoft.com/en-us/wdsi/threats/malware-encyclopedia-description?Name=VirTool%3AWin32%2FCeeInject},
  2007.

\bibitem{Renos}
{Microsoft}.
\newblock {Win32/Renos}.
\newblock
  \url{https://www.microsoft.com/en-us/wdsi/threats/malware-encyclopedia-description?name=Win32/Renos},
  2007.

\bibitem{BHO}
{Microsoft}.
\newblock {Trojan:Win32/BHO.BO}.
\newblock
  \url{https://www.microsoft.com/en-us/wdsi/threats/malware-encyclopedia-description?Name=Trojan:Win32/BHO.BO},
  2009.

\bibitem{VBinject}
{Microsoft}.
\newblock {VirTool:Win32/VBInject}.
\newblock
  \url{https://www.microsoft.com/en-us/wdsi/threats/malware-encyclopedia-description?Name=VirTool:Win32/VBInject&ThreatID=-2147367171},
  2010.

\bibitem{Vobfus}
{Microsoft}.
\newblock {Win32/Vobfus}.
\newblock
  \url{https://www.microsoft.com/en-us/wdsi/threats/malware-encyclopedia-description?name=win32%2Fvobfus},
  2010.

\bibitem{Winwebsec}
{Microsoft}.
\newblock {Win32/Winwebsec}.
\newblock
  \url{https://www.microsoft.com/en-us/wdsi/threats/malware-encyclopedia-description?Name=Win32%2fWinwebsec},
  2010.

\bibitem{Startpage}
{Microsoft}.
\newblock {Trojan:Win32/Startpage}.
\newblock
  \url{https://www.microsoft.com/en-us/wdsi/threats/malware-encyclopedia-description?Name=Trojan:Win32/Startpage},
  2011.

\bibitem{OnLineGames}
{Microsoft}.
\newblock {Win32/OnLineGames}.
\newblock
  \url{https://www.microsoft.com/en-us/wdsi/threats/malware-encyclopedia-description?name=Win32/OnLineGames},
  2015.

\bibitem{Adload}
{Microsoft}.
\newblock {Adware:Win32/Adload}.
\newblock
  \url{https://www.microsoft.com/en-us/wdsi/threats/malware-encyclopedia-description?Name=Adware:Win32/Adload},
  2018.

\bibitem{Rina_2015}
Rina Mishra and Praveen Bhanodiya.
\newblock A review on steganography and cryptography.
\newblock In {\em 2015 International Conference on Advances in Computer
  Engineering and Applications}, pages 119--122, 2015.

\bibitem{Selig_2022}
Jay Selig.
\newblock What is machine learning? a definition.
\newblock \url{https://www.expert.ai/blog/machine-learning-definition/}, 2022.

\bibitem{Stamp_2021}
Mark Stamp.
\newblock {\em Information Security: Principles and Practice}.
\newblock Wiley, 3rd edition, 2021.

\bibitem{Stamp_2022}
Mark Stamp.
\newblock {\em Introduction to Machine Learning with Applications in
  Information Security}.
\newblock Chapman and Hall/CRC, 2nd edition, 2022.

\bibitem{Stanger_2020}
James Stanger.
\newblock The ancient practice of steganography: What is it, how is it used and
  why do cybersecurity pros need to understand it.
\newblock \url{https://www.comptia.org/blog/what-is-steganography}, 2020.

\bibitem{MLP_Intro}
Hind Taud and Jean-Fran{\c{c}}ois Mas.
\newblock Multilayer perceptron ({MLP}).
\newblock In Mar{\'i}a Teresa~Camacho Olmedo, Martin Paegelow,
  Jean-Fran{\c{c}}ois Mas, and Francisco Escobar, editors, {\em Geomatic
  Approaches for Modeling Land Change Scenarios}, pages 451--455. Springer,
  2018.

\bibitem{Attention_Is_All_You_Need}
Ashish Vaswani, Noam Shazeer, Niki Parmar, Jakob Uszkoreit, Llion Jones,
  Aidan~N. Gomez, \L{}ukasz Kaiser, and Illia Polosukhin.
\newblock Attention is all you need.
\newblock In {\em Proceedings of the 31st International Conference on Neural
  Information Processing Systems}, NIPS'17, pages 6000--6010, 2017.

\bibitem{EvilModel}
Zhi Wang, Chaoge Liu, and Xiang Cui.
\newblock {EvilModel}: Hiding malware inside of neural network models.
\newblock \url{https://arxiv.org/abs/2107.08590}, 2021.

\bibitem{Wieland_2012}
A.~Wieland and C.M. Wallenburg.
\newblock Dealing with supply chain risks: Linking risk management practices
  and strategies to performance.
\newblock {\em International Journal of Physical Distribution \&\ Logistics
  Management}, 42(10):887--905, 2012.

\end{thebibliography}

\end{document}